\documentclass[aps,preprint]{revtex4}
\usepackage{graphicx}
\usepackage{lipsum}
\usepackage{float}
\usepackage{amsmath}
\usepackage{amssymb}
\usepackage{mathtools}
\include{mathmacros}

\begin{document}

\title{
Convergence towards an Erd{\H o}s-R\'enyi graph structure \\
in network contraction processes
}

\author{Ido Tishby, Ofer Biham and Eytan Katzav}

\affiliation{
Racah Institute of Physics, 
The Hebrew University, 
Jerusalem 91904, Israel
}

\begin{abstract}

In a highly influential paper twenty years ago, Barab\'asi and Albert 
[Science 286, 509 (1999)]
showed that networks undergoing generic growth processes 
with preferential attachment evolve towards scale-free structures.  
In any finite system, the growth eventually stalls and is likely to be
followed by a phase of network contraction 
due to node failures, attacks or epidemics.
Using the master equation formulation and computer simulations  
we analyze the structural evolution of 
networks subjected to contraction processes
via random, preferential 
and propagating node deletions.
We show that the contracting networks converge towards 
an Erd{\H o}s-R\'enyi network structure
whose mean degree continues to decrease as the contraction proceeds.
This is manifested by the convergence of the degree distribution towards
a Poisson distribution and the loss of degree-degree correlations. 

\end{abstract}

\maketitle

\section{Introduction}

A central observation
in contemporary science 
is that many of the processes explored take place
in complex network architectures
\cite{Havlin2010,Newman2010,Estrada2011}.
Therefore, it is of great importance to analyze the
geometries and topologies encountered in complex networks
and their temporal evolution.
Since the 1960s,
mathematical studies of networks
were focused on model systems  
such as the 
Erd{\H o}s-R\'{e}nyi (ER) network,
which exhibits a Poisson degree distribution 
with no degree-degree correlations
\cite{Erdos1959,Erdos1960,Erdos1961}.
In an ER network 
each pair of nodes is connected randomly and independently,
with equal probability
\cite{Bollobas2001}.
In fact, ER networks 
form a maximum entropy ensemble under the constraint
that the mean degree is fixed
\cite{Bauer2002,Bogacz2006,Annibale2009,Roberts2011}.
In the 1990s, the growing availability of data on large biological, social and
technological networks revolutionized the field.
Motivated by the observation that the World Wide Web 
\cite{Albert1999} 
and scientific citation networks 
\cite{Redner1998}
exhibit power-law degree distributions,
Barab\'asi and Albert (BA) introduced a simple model
that captures the essential growth dynamics of such networks
\cite{Barabasi1999}.
A key feature of the BA model is the
preferential attachment mechanism, namely the tendency of new nodes to
attach preferentially to high degree nodes.
Using mean-field equations and computer simulations
it was shown that the combination of growth and preferential attachment leads to the
emergence of scale-free networks with power-law degree distributions
\cite{Barabasi1999}. 
This result was later confirmed and generalized using more rigorous
formulations based on the master equation
\cite{Krapivsky2000,Dorogovtsev2000}
and using combinatorial methods
\cite{Bollobas2001b}.
It was subsequently found that a large variety of empirical
networks exhibit such scale-free structures,
which are remarkably different from ER networks 
\cite{Albert2002,Barabasi2009}. 

Networks are often exposed to 
node failures, attacks and epidemics, 
which may halt their growth
and lead to their contraction and eventual collapse.
Since network growth and contraction are kinetic nonequilibrium processes  
they are irreversible, namely the contraction process 
is not the same as the growth process played backwards in time.
This hysteretic behavior is analogous to the response of magnetic systems 
to an external magnetic field, where the magnetization depends not only on the 
instantaneous field but also on its history.

One can distinguish between three generic scenarios of network contraction:
the {\it random deletion} scenario that describes 
inadvertent random failures of nodes
\cite{Cohen2000}, the 
{\it preferential deletion} scenario that describes intentional attacks
\cite{Cohen2001}, which
are more likely to focus on high-degree nodes,
and the {\it propagating deletion} scenario that describes cascading failures 
that spread throughout the network 
\cite{Crucitti2004,Buldyrev2010,Satorras2015}. 
Using the framework of percolation theory, 
it was shown that in the final stages of the contraction process
the network breaks down into disconnected components
\cite{Albert2000,Cohen2000,Cohen2001,Braunstein2016}.
However, the evolution of the network structure 
throughout the contraction phase has not been studied
in a systematic way.

In this paper we analyze the structural evolution of networks during the contraction process.
To this end we derive a master equation for
the time dependence of the degree distribution
during network contraction via the random, preferential and propagating
node deletion scenarios.
We show that the Poisson distribution with a time dependent mean
degree $c_t$ is a solution of the master equation.
Moreover, using the relative entropy between the degree distribution $P_t(k)$
of the contracting network at time $t$
and the corresponding Poisson distribution $\pi_t(k)$ with the same 
mean degree 
$c_t=\langle K \rangle_t$ 
we show that
the Poisson distribution is an attractive solution
for the degree distributions of random networks that contract via
these three network contraction scenarios.
Thus, the degree distribution $P_t(k)$ converges towards
$\pi_t(k)$ during the contraction process.
Using computer simulations of contracting networks we show that
in case that the initial network exhibits degree-degree correlations
these correlations decay during the contraction process.
We thus conclude that the contracting networks converge towards
an ER structure whose mean degree continues to decrease as
the contraction proceeds.

The paper is organized as follows. 
In Sec. II we describe the three network contraction scenarios considered
in this paper and discuss related examples of contraction processes in empirical networks.
In Sec. III we present the master equation for these three  
network contraction scenarios.
In Sec. IV we show that a Poisson distribution with
a time dependent mean degree is a stationary solution of the
master equation. 
In Sec. V we use direct integration of the master equation
in conjunction with computer simulations to
examine the convergence of the degree distribution of a contracting 
network towards a Poisson distribution.
In Sec. VI we use the relative entropy $S_t$ between the 
degree distribution $P_t(k)$ of the contracting network at time $t$ and
the corresponding Poisson distribution $\pi_t(k)$ with the 
same mean degree $c_t=\langle K \rangle_t$ to quantify the
rate of convergence of $P_t(k)$ towards $\pi_t(k)$.
In Sec. VII we use computer simulations to evaluate the decay rate
of the degree-degree correlation function
$\Gamma_t$ during the contraction process. 
The results are discussed in Sec. VIII and summarized in Sec. IX.
In Appendix A we present a detailed derivation of the master equation
for the three network contraction scenarios. 
In Appendix B we present an exact solution for the time dependent 
degree distribution $P_t(k)$ of a network contracting via the random
node deletion scenario.

\section{Network contraction processes}

We consider network contraction processes in which at each time
step a single node is deleted together with its links.
The size of the network at time $t$ is thus $N_t = N - t$,
where $N_0=N$ is the size of the initial network.
Consider a node of degree $k$, whose neighbors are of
degrees $k'_r$, $r=1,2,\dots,k$.
Upon deletion of such node 
the degrees of its neighbors are reduced to
$k'_r-1$, $r=1,2,\dots,k$. 
The node deleted at each time step is selected randomly.
However, the probability of a specific node to be deleted 
in the next time step may depend on
its degree as well as on other properties, according to the specific network
contraction scenario.
Here we focus on three generic scenarios of network contraction:
the random node deletion scenario that describes the
random, inadvertent failure of nodes, the 
preferential node deletion scenario that describes intentional attacks that
are more likely to focus on highly connected nodes
and the propagating node deletion scenario that describes cascading failures 
that spread throughout the network. 

In the random deletion scenario, at each time step a random 
node is selected for deletion.
In this scenario each one of the nodes in the network 
at time $t$ has the same probability to be
selected for deletion, regardless of its degree or any other properties.
Since at time $t$ there are $N_t$ nodes in the network,
the probability of each one of them to be selected for deletion is $1/N_t$.
This scenario may describe a situation in which random nodes in 
a communication network become dysfunctional independently of each other due to
technical failures or random attacks
\cite{Albert2000,Cohen2000}.

In the preferential deletion scenario  
the probability of a node to be targeted for deletion at a given time 
step is proportional to its degree.
This means that the probability of a given node of degree $k$ to be 
deleted at time $t$ is $k/[N_t \langle K \rangle_t]$.
This is equivalent to picking a random edge in the network and 
randomly selecting for deletion one of the two nodes at its ends.
This scenario may describe attacks in which high degree nodes
are more likely to be targeted
\cite{Cohen2001}.

In the propagating deletion scenario at each time
step the node to be deleted is randomly selected among the
neighbors of the node deleted in the previous time step.
In case that the node deleted in the previous time step does
not have any yet-undeleted neighbor we pick a random node,
randomly select one of its neighbors for deletion and continue
the process from there.
This scenario may describe cascading failures in which the
failure of a node increases the load on its neighbors 
and causing their subsequent failure.
Such situations may occur in power grids and transportation networks
\cite{Daqing2014,Schafer2018}.
Another mechanism of cascading failures was identified in social
networks in which a user who leaves the network may encourage 
some of his/her friends to leave the network too, possibly for joining
a competing network 
\cite{Torok2017,Lorincz2019}.

\section{The master equation}

Consider an ensemble of 
networks of size $N_0$ at time $t=0$,
with degree distribution $P_0(k)$ and
mean degree $\langle K \rangle_0$,
which are exposed to network contraction via node deletion. 
Below we derive a master equation that describes the time
evolution of the degree distribution $P_t(k)$ throughout the
contraction process. The master equation consists of a set of
coupled first-order differential equations of the form
\cite{vanKampen2007,Gardiner2004}

\begin{equation}
\frac{d}{dt} \vec P_t = M \vec P_t,
\label{eq:M}
\end{equation}

\noindent
where 
$\vec P_t$ is a vector whose elements are the 
probabilities $P_t(k)$, $k=0,1,2,\dots$, and $M$ is the transition matrix.

At each time step during the contraction process 
a single node is deleted from the network.
In addition to the primary effect of the loss of the deleted node,
the network sustains a secondary effect as 
the neighbors of the deleted node lose one link each.
An intrinsic property of the secondary effect is that it is
of a preferential nature, namely the likelihood of a node
to be a neighbor of the deleted node is proportional to its degree.
The time dependent degree distribution is given by

\begin{equation}
P_t(k) = \frac{N_t(k)}{N_t},
\label{eq:P_t(k)}
\end{equation}

\noindent
where $N_t(k)$ is the number of nodes of degree $k$ at time $t$.
The mean degree of the contracting network at time $t$ is
given by

\begin{equation}
\langle K \rangle_t = \sum_{k=0}^{\infty} k P_t(k),
\end{equation}

\noindent
while the second moment of the degree distribution is
given by

\begin{equation}
\langle K^2 \rangle_t = \sum_{k=0}^{\infty} k^2 P_t(k).
\end{equation}

Here we analyze three generic scenarios of network contraction:
the random deletion scenario, in which a randomly
selected node is deleted at each time step;
the preferential deletion scenario  
in which the probability of a node to
be targeted for deletion is proportional to its degree;
and the propagating deletion scenario in which at each time
step we delete a random neighbor of the last deleted node.
To demonstrate the derivation of the master equation
we consider below the relatively simple case of random node deletion
(for a detailed derivation of the master equation
for all three network contraction scenarios see Appendix A).
The time dependence of $N_t(k)$
depends on the primary effect, given by the probability that the node selected
for deletion is of degree $k$, as well as on the secondary effect  of node deletion
on neighboring nodes of degrees $k$ and $k+1$.
In random node deletion the probability that the node 
selected for deletion at time $t$ is of degree $k$ is
given by $N_t(k)/N_t$.
Thus, the rate in which $N_t(k)$ decreases due to the primary effect of 
the deletion of nodes of degree $k$  
is given by

\begin{equation}
R_t(k \rightarrow \varnothing) = \frac{N_t(k)}{N_t},
\label{eq:Rk}
\end{equation}

\noindent
where $\varnothing$ represents the empty set.
In case that the node deleted at time $t$ is of degree $k'$,
it affects $k'$ adjacent nodes, which lose one link each. 
The probability of each one of these $k'$ nodes
to be of degree $k$ is given by
$k N_t(k)/[ N_t \langle K\rangle_t ]$.
We denote by $W_t(k \rightarrow k-1)$ the expectation value of
the number of nodes of degree $k$ that lose a link at time $t$ and
are reduced to degree $k-1$.
Summing up over all possible values of $k'$,  
we find that the secondary effect of random node deletion on nodes of 
degree $k$ accounts to

\begin{equation}
W_t(k \rightarrow k-1) =  \frac{kN_t(k)}{N_t}.
\label{eq:Wk}
\end{equation}

\noindent
Similarly, the secondary effect on nodes of degree $k+1$ 
accounts to

\begin{equation}
W_t(k+1 \rightarrow k) =  \frac{ (k+1)N_t(k+1)}{N_t}.
\label{eq:Wk+1}
\end{equation}

\noindent
The time evolution of $N_t(k)$ can be expressed in terms
of the forward difference

\begin{equation}
\Delta_t N_t(k) = N_{t+1}(k) - N_t(k).
\end{equation}

\noindent
Combining the primary and the
secondary effects on the time dependence of $N_t(k)$ 
we obtain

\begin{equation}
\Delta_t N_t(k) =
- R_t(k \rightarrow \varnothing) + \left[ W_t(k+1 \rightarrow k) - W_t(k \rightarrow k-1) \right].
\label{eq:RWW}
\end{equation}

\noindent
Inserting the expressions for $R_t(k \rightarrow \varnothing)$, $W_t(k \rightarrow k-1)$ and
$W_t(k+1 \rightarrow k)$ from Eqs.
(\ref{eq:Rk}), (\ref{eq:Wk}) and (\ref{eq:Wk+1}), respectively,
we obtain

\begin{equation}
\Delta_t N_t(k) =
\frac{(k+1)[ N_t(k+1) - N_t(k) ]}{N_t}.
\label{eq:DeltaNtk}
\end{equation}

\noindent
Since nodes are discrete entities the process of node deletion
is intrinsically discrete. Therefore, the replacement of the forward difference
$\Delta_t N_t(k)$
by a time derivative of the form 
$d N_t(k)/dt$ involves an approximation.
In fact, it is closely related to the approximation made in numerical
integration of differential equations using the Euler method
\cite{Butcher2003}.
In the Euler method the time derivative $df_t/dt$ is replaced by
$(f_{t+h}-f_t)/h$, where $h$  is a
suitably chosen time step.
In our case $h= 1$. 
Below we evaluate the error associated with this approximation.
To this end we use a series expansion of the form

\begin{equation}
\Delta_t N_t(k) =
\frac{d}{dt} N_t(k) + \frac{1}{2} \frac{d^2}{dt^2} N_t(k)
+ \dots.
\end{equation}

\noindent
Typical degree distributions, which are not too narrow,
satisfy $N_t(k) \ll N_t$ for any value of $k$.
For such distributions
the second time derivative satisfies

\begin{equation}
\frac{1}{2} \frac{d^2}{dt^2} N_t(k) \sim O \left( \frac{1}{N_t^2} \right),
\end{equation}

\noindent
and quickly vanishes for sufficiently large networks.
This means that the replacement of the forward difference by
a time derivative has little effect on the results.
Thus, the difference equation (\ref{eq:DeltaNtk}) can be replaced by the
differential equation

\begin{equation}
\frac{d}{dt} N_t(k) =
\frac{(k+1)[ N_t(k+1) -  N_t(k) ]}{N_t} + O \left( \frac{1}{N_t^2} \right).
\label{eq:DeltaNtkde}
\end{equation}

\noindent
In a more rigorous approach one could define
a reduced time $\theta = t/N$ and a density $n(\theta,k) = N_t(k)/N$,
as done in Refs. 
\cite{Kurtz1970,Kurtz1981,Wormald1995,Wormald1999,Warnke2019}.
Using this approach, one can show that the random variable
$N_{t=\theta N}(k)/N$ concentrates, in the large $N$ limit, around the
deterministic density $n(\theta,k)$ which is the solution of the
corresponding differential equation.

The derivation of the master equation is completed by taking the
time derivative of Eq. (\ref{eq:P_t(k)}), which is given by

\begin{equation}
\frac{d}{dt} P_t(k) = \frac{1}{N_t} \frac{d}{dt} N_t(k) - \frac{N_t(k)}{N_t^2} \frac{d}{dt} N_t.
\label{eq:dPt_Nt}
\end{equation}

\noindent
Inserting the time derivative of $N_t(k)$ from Eq. (\ref{eq:DeltaNtkde}),
and using the fact that
$d N_t/dt=-1$,
we obtain the master equation for the random deletion scenario,
which is given by

\begin{equation}
\frac{d}{dt} P_t(k)=
\frac{1}{N_t}
\left[ (k+1)P_t(k+1) - k P_t(k) \right].
\label{eq:dP(t)/dtRC0}
\end{equation}

\noindent
The derivation of the master equations for the preferential deletion 
and the propagating deletion scenarios can be performed along similar lines.
The detailed derivations of the master equations for all three scenarios appear in Appendix A.
Interestingly, the resulting master equations for these three network contraction scenarios
can be written in a unified manner, in the form

\begin{equation}
\frac{d}{dt} P_t(k) 
= 
\frac{A_t}{N_t}
\left[ (k+1) P_t(k+1) - k P_t(k) \right]
- 
\frac{B_t(k)}{N_t}
P_t(k),
\label{eq:dP/dt}
\end{equation}

\noindent
where the coefficients are given by

\begin{equation}
A_t = 
\left\{
\begin{array}{ll}
  1   &  {\rm \ \ \ \ \   random \ deletion} \\
  \frac{\langle K^2 \rangle_t }{\langle K \rangle_t^2}   
& {\rm \ \ \ \ \   preferential \ deletion}  \\
      \frac{\langle K^2 \rangle_t - 2 \langle K \rangle_t}{\langle K \rangle_t^2} & 
{\rm \ \ \ \ \  propagating \ deletion}
\end{array}
\right.
\end{equation}

\noindent
and

\begin{equation}
B_t(k) =
\left\{
\begin{array}{ll}
 0  &  {\rm \ \ \ \ \  random \ deletion}    \\
  \frac{k - \langle K \rangle_t}{\langle K \rangle_t}  
&  {\rm \ \ \ \ \    preferential \ deletion}    \\
  \frac{k - \langle K \rangle_t}{\langle K \rangle_t}  
& {\rm \ \ \ \ \   propagating \ deletion}.
\end{array}
\right.
\end{equation}

\noindent
The master equation consists of a set of coupled ordinary differential equations
for $P_t(k)$, $k=0,1,2,\dots,k_{\rm max}$.
In order to calculate the time evolution of the degree distribution $P_t(k)$ during
the contraction process one solves the master equation using direct numerical integration,
starting from the initial network that consists of $N_0$ nodes whose degree distribution
is $P_0(k)$. For any finite network the degree distribution is bounded from above
by an upper bound denoted by $k_{\rm max}$, which satisfies
the condition $k_{\rm max} \le N_0-1$. Since the contraction process can only
delete edges from the remaining nodes and cannot increase the degree of any
node, the upper cutoff $k_{\rm max}$ is maintained throughout the contraction
process. 

Expressing the master equation in terms of the transition rate matrix formulation of Eq. (\ref{eq:M}),
it is found that the matrix $M$ is an upper bidiagonal matrix, whose diagonal
elements are given by

\begin{equation}
M_{k,k} = - \frac{k A_t + B_t(k)}{N_t},
\end{equation}

\noindent
the off-diagonal elements are given by

\begin{equation}
M_{k,k+1} = \frac{(k+1) A_t}{N_t},
\end{equation}

\noindent
and $M_{k,k'}=0$ for $k'<k$ and $k'>k+1$.

The rate coefficients on the right hand side of the master equation (\ref{eq:dP/dt})
include a combination of explicit and implicit time dependence. The overall factor of
$1/N_t$ is the only components that exhibits an explicit time dependence,
while the moments $\langle K \rangle_t$ and $\langle K^2 \rangle_t$
depend implicitly on the time via the instantaneous degree distribution $P_t(k)$.
Since their coefficients are time dependent they need to be updated
throughout the numerical integration of Eq. (\ref{eq:dP/dt}).
In particular, the instantaneous network size $N_t$ should be updated
at each time step. The time derivatives of the moments
$\langle K \rangle_t$ and $\langle K^2 \rangle_t$
scale with the network size like $1/N_t$. Therefore, they may
be considered as slow variables and updated once every several 
time steps during the integration of the master equation.

Since the only explicit time dependence of the rate coefficients
on the right hand side of Eq. (\ref{eq:dP/dt}) is via the overall
factor of $1/N_t$ one can multiply both sides of the equation by
$N_t$. The time derivative on the left hand side 
of Eq. (\ref{eq:dP/dt})
can then be
expressed in terms of $d \tau = dt/N_t$.
This implies that the time dependence of $P_t(k)$ 
is expressed in terms of the ratio $N_t/N_0$,
or more specifically in terms of $\tau = \ln (N_t/N_0)$. 
This means that the initial network size is essentially an
extensive parameter while the time is measured in terms of
the fraction of the network that remains.
This conclusion is of great practical importance because it means
that for any given degree distribution it is sufficient to perform the
simulation of network collapse for one size of the initial network.

The first term on the right hand side of Eq. (\ref{eq:dP/dt}) 
is referred to as the trickle-down term
\cite{TrickleDown}.
This term represents
the step by step downwards flow of probability from high to low degrees.  
The coefficient $A_t$ of the trickle-down term depends on the network 
contraction scenario.
In random deletion $A_t=1$,
because the probability of a node to be selected for
deletion does not depend on its degree.
In preferential deletion $A_t$ is proportional to $\langle K^2 \rangle_t$
because the probability of a node to be deleted is proportional
to its degree $k$ while the magnitude of the secondary effect is also proportional to $k$.

The second term on the right hand side of Eq. (\ref{eq:dP/dt}) 
is referred to as the redistribution term.
This term vanishes in the random deletion scenario.
However, in the preferential and propagating deletion scenarios
the redistribution term 
is negative for 
$k > \langle K \rangle_t$
and positive for
$k < \langle K \rangle_t$.
Thus the redistribution term decreases the probabilities
$P_t(k)$ for values of $k$ that are above the mean degree
and increases them for values of $k$ that are below the mean 
degree. 
Moreover, in absolute value the size of the redistribution term is proportional to
$|k - \langle K \rangle_t|$,
which means that nodes of degrees that are much higher 
or much lower than $\langle K \rangle_t$ are most strongly affected
by this term.

In general, the master equation accounts for the time evolution of
the degree distribution over an ensemble of networks of the same
initial size $N_0$ and degree distribution $P_0(k)$, which are exposed
to the same network contraction scenario.
A fundamental question in this context is to what extent the solution
of a deterministic differential equation describes the results of single
instances of the stochastic process in systems of finite size.
In the context of network contraction processes, a single instance
of the stochastic process at time $t$ is described by $N_t(k)$, $k=0,1,\dots$.
The corresponding results of the master equation are given by
$\langle N_t(k) \rangle = N_t P_t(k)$, $k=0,1,\dots$.
Using the theory of stochastic processes it was shown that under
very general conditions the results of single instances,
given by $N_t(k)$ are narrowly distributed around 
$\langle N_t(k) \rangle$, thus the master equation provides
a good description of the corresponding stochastic process
\cite{Kurtz1970,Kurtz1981,Wormald1995,Wormald1999,Warnke2019}.

\section{The Poisson solution}

Consider an ER network of $N_t$ nodes with mean degree 
$c_t = \langle K \rangle_t$.
Its degree distribution follows
a Poisson distribution of the form

\begin{equation}
\pi_t(k) = \frac{ e^{-c_t} c_t^k }{k!}.
\label{eq:poisson}
\end{equation}

\noindent
The second moment of the degree distribution satisfies
$\langle K^2 \rangle_t = c_t(c_t + 1)$.
To examine the contraction process of ER networks
we start from an initial network of $N_0$ nodes
whose degree distribution follows a Poisson distribution $\pi_0(k)$
with mean degree $c_0$.
Inserting 
$\pi_t(k)$
into the master equation (\ref{eq:dP/dt}) we find that 
the time derivative on the left hand side is given by

\begin{equation}
\frac{d}{dt} \pi_t(k) = 
- \frac{d c_t}{dt}
\left( 1 - \frac{k}{c_t} \right)  \pi_t(k),
\label{eq:dpi/dt1}
\end{equation}

\noindent
On the other hand,
inserting $\pi_t(k)$ on the right hand side of Eq. (\ref{eq:dP/dt}),
we obtain

\begin{equation}
\frac{d}{dt} \pi_t(k) =
\frac{A_t}{N_t} (c_t - k) \pi_t(k) - \frac{B_t(k)}{N_t} \pi_t(k),
\label{eq:dpi/dt2}
\end{equation}

\noindent
where

\begin{equation}
A_t = 
\left\{
\begin{array}{ll}
  1   &  {\rm \ \ \ \ \   random \ deletion} \\
  \frac{c_t + 1}{c_t}   
& {\rm \ \ \ \ \   preferential \ deletion}  \\
      \frac{c_t - 1}{c_t} & 
{\rm \ \ \ \ \  propagating \ deletion}
\end{array}
\right.
\end{equation}

\noindent
and

\begin{equation}
B_t(k) =
\left\{
\begin{array}{ll}
 0  &  {\rm \ \ \ \ \  random \ deletion}    \\
  \frac{k - c_t}{c_t}  
&  {\rm \ \ \ \ \    preferential \ deletion}    \\
  \frac{k - c_t}{c_t}  
& {\rm \ \ \ \ \   propagating \ deletion}.
\end{array}
\right.
\end{equation}

\noindent
In order that $\pi_t(k)$ will be a solution of Eq. (\ref{eq:dP/dt}),
the right hand sides of Eqs. (\ref{eq:dpi/dt1}) and (\ref{eq:dpi/dt2}) 
must coincide.
In the case of random deletion this implies that

\begin{equation}
\frac{1}{c_t} \frac{d c_t}{dt} = - \frac{1}{N_t}.
\end{equation}

\noindent
Integrating both sides for $t'=0$ to $t$, we obtain
the condition 

\begin{equation}
c_t = c_0 \frac{N_t}{N_0} = c_0 - \frac{c_0}{N_0} t.
\end{equation}

\noindent
Repeating the analysis presented above for the
cases of preferential deletion and propagating deletion
it is found that $\pi_t(k)$
solves the master equation (\ref{eq:dP/dt})
for the three network contraction scenarios,
while
the mean 
degree, $c_t$ decreases linearly in time
according to

\begin{equation}
c_t = c_0 - R t,
\label{eq:c_linear}
\end{equation}

\noindent
where
the rate $R$ is given by

\begin{equation}
R = 
\left\{
\begin{array}{ll}
 \frac{c_0}{N_0}   & {\rm \ \ \ \ \  random \ deletion}   \\
 \frac{c_0+2}{N_0}   &  {\rm \ \ \ \ \  preferential \ deletion}  \\
 \frac{c_0}{N_0}   & {\rm \ \ \ \ \  propagating \ deletion}.
\end{array}
\right.
\label{eq:c_t}
\end{equation}

\noindent
This means that an ER network exposed to
any one of the three contraction scenarios
remains an ER network at all times,
with a mean degree that decreases according to Eq.
(\ref{eq:c_linear}). The network size at time $t$ is
$N_t = N_0 - t$, where $N_0$ is the initial size.

In the case of random deletion the contraction process ends
at time $t=N_0$ when the network vanishes completely.
In the case of preferential deletion the deleted node at each
time step is picked via a randomly selected edge. Therefore,
once a node becomes isolated it will never be selected for
deletion. As a result, the process of preferential deleted comes
to a halt once all the remaining nodes become isolated and $c_t=0$.
Using Eqs. (\ref{eq:c_linear}) and (\ref{eq:c_t}) we find that
this happens at time $t_{\rm h} = c_0 N_0/(c_0+2)$.
Thus, the number of isolated nodes that remain is
$N_{\rm h} = 2 N_0/(c_0+2)$.
In the case of propagating deletion one may encounter a
situation in which the node deleted at time $t$ becomes isolated,
namely it does not have any yet-undeleted neighbors.
In this case we continue the deletion process by selecting
a random node, randomly selecting one of its neighbors
for deletion and continuing the process from there.

\section{Numerical integration and computer simulations}

To test the convergence of contracting networks
towards the ER structure we study the three network
contraction scenarios presented above using numerical
integration of the master equation and computer simulations.
As an initial network we use the BA network, which is a
scale-free networks with a
power-law degree distribution of the form 
$P_0(k) \sim k^{-\gamma}$,
where 
$\gamma=3$
\cite{Barabasi1999,Krapivsky2000,Dorogovtsev2000,Bollobas2001}.
To generate the initial networks for the computer simulations we use
the BA growth model, in which at each time step a new
node is added to the network and is connected preferentially 
by undirected edges to $m$ of the existing nodes
\cite{Barabasi1999}.
The $m$ edges of the new node are added sequentially, under the
condition that each existing node can receive at most one of these 
edges (multiple edges are not allowed, thus the resulting network is a simple graph).
The preferential attachment property implies that the 
probability of an existing node whose degree at time $t$ is 
$k$ (and is not yet connected to the new node) to receive the
next link from the new node is proportional to $k$.
The parameter $m$ may take any nonzero integer value. 
In the special case of $m=1$
the resulting network exhibits a tree structure while for 
$m \ge 2$ it includes cycles.
As a seed network for the growth process we use a complete graph of $m+1$ nodes,
such that at time $t=0$ all the nodes in the seed network are of degree $m$.
Since there are $m$ edges that are added to the network with each new node,
and each edge connects two nodes, in the large network limit $N_0 \gg m$,
the mean degree is $\langle K \rangle_0 = 2m$.
Thus, the network becomes more dense as $m$ is increased.
Since the seed network consists of a single connected component, the resulting
network consists of a single component at all times.
The growth process ends when the network reaches the desired size,
denoted by $N_0$.
The degree distribution of a BA network is given by

\begin{equation}
P_0(k) \sim k^{-\gamma}, 
\label{eq:P0k}
\end{equation}

\noindent
where $\gamma=3$.
Since the degree of each new node upon formation is $m$,
the lower bound of the degree distribution
(\ref{eq:P0k})
is $k_{\rm min}=m$.
To generate the initial degree distribution used in the direct integration
of the master equation we use the master equation that describes the
BA network growth process
\cite{Krapivsky2000,Dorogovtsev2000},
with the same seed network used in the computer simulations.

\begin{figure}
\begin{center}
\includegraphics[width=13.5cm]{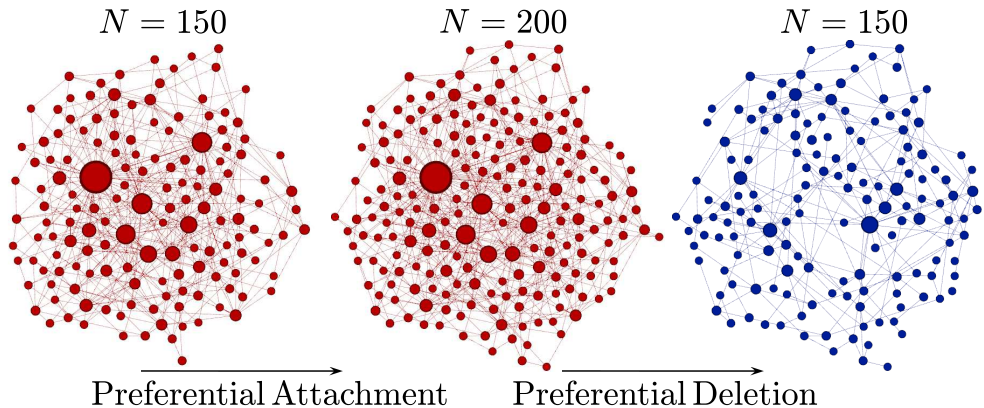} 
\caption{
(Color online)
A BA network is shown during the growth phase,
at sizes of $N=150$ (left)
and $N=200$ (middle), and during the contraction phase when its size is reduced to
$N=150$ (right). There is a striking difference between the
structures of the growing network that exhibits large hubs 
and the contracting network that shows little variation between the degrees
of different nodes. 
In a supplemental movie \cite{SM} we present the full evolution of this network 
during the growth phase and the subsequent contraction phase via random 
deletion and preferential deletion.
}
\label{fig:1}
\end{center}
\end{figure}

In Fig. \ref{fig:1} we present the structure
of a BA network with $m=3$ 
during growth at an 
intermediate size of $N=150$ (left) and at the final size of $N=200$ (middle).
At this point
the network starts to 
contract via preferential node deletion.
The structure of the network during the contraction process is presented (right),
when its size is down to $N=150$.
To emphasize the variation in the degrees of different nodes, each node
is represented by a circle whose area is proportional to the degree of the node.
It is apparent that the initial BA network includes several dominant
hubs, as expected in a scale-free network, while in the network depicted during  
contraction there is little variation between the degrees of different nodes.
In the supplemental movie 
\cite{SM}
we present the evolution
of the structure of the same BA network instance during the growth phase
and the subsequent contraction phase via random deletion and preferential
deletion. 

\begin{figure}
\begin{center}
\includegraphics[width=15.6cm]{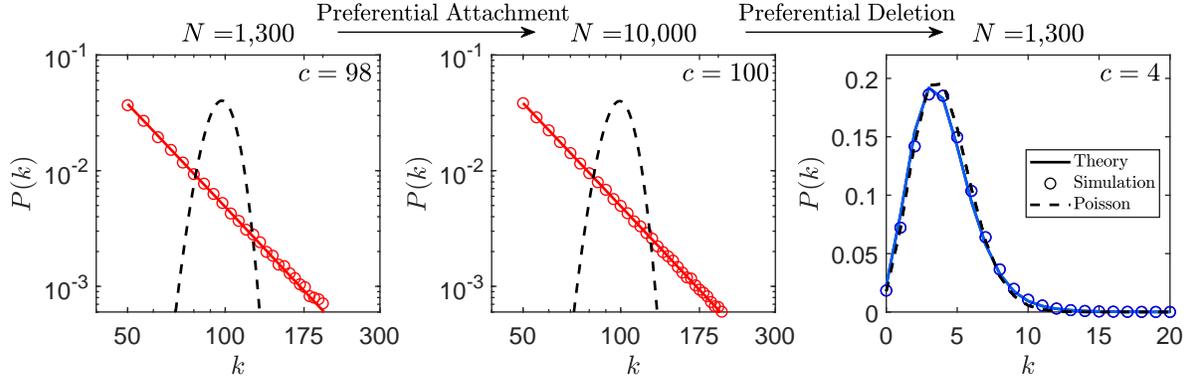}
\caption{
(Color online)
The degree distributions $P(k)$ of a BA network during growth,
obtained from numerical integration of the master equation of 
Refs. 
\cite{Krapivsky2000} 
and 
\cite{Dorogovtsev2000}
(solid line)
and from computer simulations (circles) at an
intermediate size of $N=1,300$ (left) and at
the final size of $N=10,000$ (middle).
The Poisson distribution with the same mean degree is also shown (dashed line).
At $N=10,000$ the network starts to contract via preferential node deletion.
The degree distribution $P(k)$ of the contracted network is shown (right) 
when it is reduced back to $N=1,300$ nodes.
The theoretical results (solid line) obtained from the master equation 
[Eq. (\ref{eq:dP/dt})] are in very good agreement with computer simulations
(circles) and with the Poisson distribution with the same mean degree (dashed line).
}
\label{fig:2}
\end{center}
\end{figure}

In Fig. 
\ref{fig:2}
we present
the degree distributions $P(k)$ 
(solid lines)
of a BA network 
with $m=50$,
obtained from numerical integration of the master equation
that describes the growth process
\cite{Krapivsky2000,Dorogovtsev2000}
during growth at an intermediate size of
$N=1,300$ (left) and at the final size of 
$N=10,000$ (middle).
The resulting degree distributions,
presented in a log-log scale,
follow a straight line that corresponds to
$P(k) \sim k^{-\gamma}$,
with $\gamma=3$.
They coincide with the
degree distributions obtained from computer
simulations of the BA growth process (circles).
The corresponding Poisson distributions with the same
value of the mean degrees, namely $c=\langle K \rangle$,
are also shown (dashed lines).
They form narrow and nearly symmetric distributions whose
peaks are close to the mean degree $c$.
Clearly, the power-law distribution (solid line) and the Poisson
distribution (dashed line) are essentially as different from each
other as any two distributions with the same mean degree could be.
Starting from $N=10,000$ the
network contracts via the preferential node deletion scenario.
The degree distribution of the contracted network 
when its size is reduced to $N=1,300$ nodes is shown (right).
The results obtained from numerical integration of the master equation
(\ref{eq:dP/dt}) and from computer simulations (solid line and circles, respectively)
are found to be in excellent agreement with the corresponding Poisson distribution 
with the same mean degree (dashed line).

\begin{figure}
\begin{center}
\includegraphics[width=5.4cm]{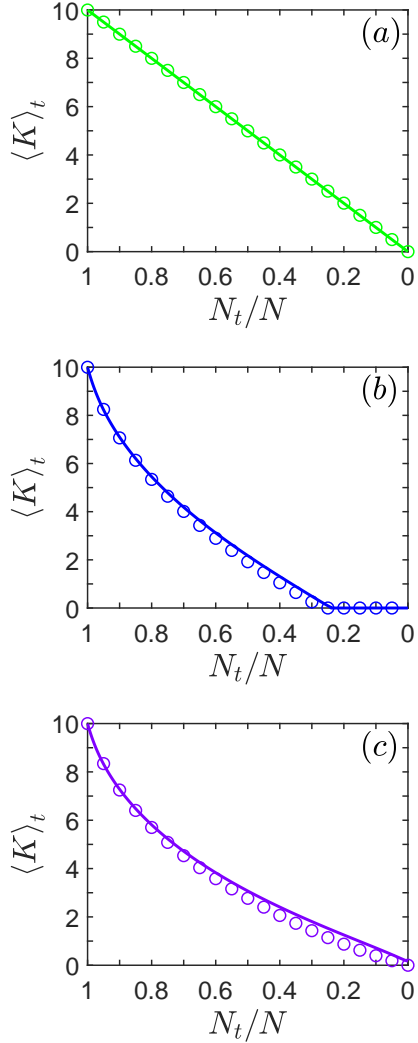}
\caption{
(Color online)
The mean degrees  
$\langle K \rangle_t$ vs. $N_t/N$, 
obtained from numerical integration of the master equation 
(solid lines), for networks that contract via
random deletion (a), preferential deletion (b) 
and propagating deletion (c),
starting from a BA network with $m=5$ of size $N=10,000$.
The master equation results are in very good agreement 
with computer simulation results (circles).
In the case of random node deletion $\langle K \rangle_t$ decreases
linearly in time according to
$\langle K \rangle_t = \langle K \rangle_0 - R t$,
where $R = c_0/N$ is independent of the degree distribution
of the initial network.
In the preferential deletion and propagating deletion scenarios
the time dependence of $\langle K \rangle_t$ during the contraction
process depends on the degree distribution of the initial network.
In case that the initial network exhibits a power-law distribution it
is found that in the early stages $\langle K \rangle_t$ quickly decreases
due to the preferential deletion of high degree nodes. 
The decay rate of $\langle K \rangle_t$ gradually slows down and
approaches a constant slope as $P_t(k)$ converges towards
a Poisson distribution.
}
\label{fig:3}
\end{center}
\end{figure}

In Fig. \ref{fig:3} we present the evolution of the mean
degree $\langle K \rangle_t$ as a function of time  
for random deletion (a),
preferential deletion (b)
and propagating deletion (c).
In the random deletion scenario, the mean degree
$\langle K \rangle_t$ decreases linearly in time,
where
$\langle K \rangle_t = \langle K \rangle_0 ( 1 - {t}/{N_0} )$
does not depend on the functional form of $P_0(k)$.
In the preferential and propagating deletion scenarios
the decay rate of $\langle K \rangle_t$ depends on the
initial degree distribution $P_0(k)$.
In case that $P_0(k)$ is a fat tailed distribution
such as the power-law distribution
the initial decay of $\langle K \rangle_t$ is fast and then it slows down.
This is due to the fact that in these two scenarios
an excess of high degree nodes are targeted for deletion 
in the early stages, enhancing the decrease of 
$\langle K \rangle_t$.

\section{The relative entropy}

In order to establish that networks exposed to these contraction scenarios actually converge
towards the ER structure, it remains to show that this asymptotic solution is attractive.
To this end we 
quantify the convergence rate of $P_t(k)$ towards the Poisson distribution,
using the relative entropy
(also referred to as the Kullback-Leibler divergence),
defined by
\cite{Kullback1951}

\begin{equation}
S_t =
\sum_{k=0}^{\infty} P_t(k)
\ln \left[ \frac{P_t(k)}{\pi_t(k)} \right],
\label{eq:S}
\end{equation}

\noindent
where $\pi_t(k)$ is the Poisson distribution given by Eq. (\ref{eq:poisson}).
The relative entropy $S_t$ measures the difference between
the probability distribution $P_t(k)$ and the reference distribution $\pi_t(k)$.
It also quantifies the added information associated with constraining the degree
distribution $P_t(k)$ rather than only the mean degree $c_t$
\cite{Annibale2009,Roberts2011}.
The Poisson distribution is a proper reference distribution because it satisfies
$\pi_t(k) > 0$ for all the non-negative integer values of $k$.
The relative entropy is always non-negative and satisfies 
$S_t=0$ if and only if $P_t(k) = \pi_t(k)$.
Therefore, $S_t$ can be used as
a measure of the distance between a given network and
the corresponding ER network with the same mean degree.
In each of the network contraction processes,
the degree distribution $P_t(k)$ evolves in time according to 
Eq. (\ref{eq:dP/dt}).
As a result, the relative entropy $S_t$ of the network also evolves as the
network contracts.
The time derivative of $S_t$ is given by

\begin{equation}
\frac{d}{dt} S_t =
\sum_{k=0}^{\infty} 
\ln \left[ \frac{P_t (k)}{\pi_t(k)} \right] 
\frac{d}{dt}P_t(k)
+
\sum_{k=0}^{\infty} 
\frac{d}{dt} P_t(k) 
-
\sum_{k=0}^{\infty}
\frac{P_t(k)}{\pi_t(k)}
\frac{d}{dt} \pi_t(k).
\label{eq:ds/dt_full}
\end{equation}

\noindent
Replacing the order of the summation and the derivative in
the second term on the right hand side of Eq.
(\ref{eq:ds/dt_full}),
we obtain

\begin{equation}
\sum_{k=0}^{\infty} 
\frac{d}{dt} P_t(k) =
\frac{d}{dt} \left[ \sum_{k=0}^{\infty} P_t(k) \right] =0.
\end{equation}

\noindent
Inserting the derivative $d \pi_t(k)/dt$ from Eq. (\ref{eq:dpi/dt1})
into the third term on the right hand side of
Eq. (\ref{eq:ds/dt_full}),
and recalling that $c_t = \langle K \rangle_t$,
we obtain

\begin{equation}
\sum_{k=0}^{\infty}
\frac{P_t(k)}{\pi_t(k)}
\frac{d}{dt} \pi_t(k) =
-\frac{dc_t}{dt}
\sum_{k=0}^{\infty}
\left( 1-\frac{k}{c_t} \right) P_t(k) = 0.
\end{equation}

\noindent
Since the second and third terms in
Eq. (\ref{eq:ds/dt_full}) vanish,
the time derivative of the relative entropy is given by

\begin{equation}
\frac{d}{dt} S_t
=
\sum_{k=0}^{\infty} 
\ln \left[ \frac{P_t(k)}{\pi_t(k)} \right]
\frac{d}{dt} P_t(k).
\label{eq:ds/dt}
\end{equation}

\noindent
This is a general equation that applies to any network contraction scenario
in which the Poisson distribution $\pi_t(k)$ is a solution.
In order to obtain a more specific equation for a given
network contraction scenario, one should insert the expression
for the derivative $dP_t(k)/dt$ from the corresponding master equation
into Eq. (\ref{eq:ds/dt}).

\begin{figure}
\begin{center}
\includegraphics[width=5.4cm]{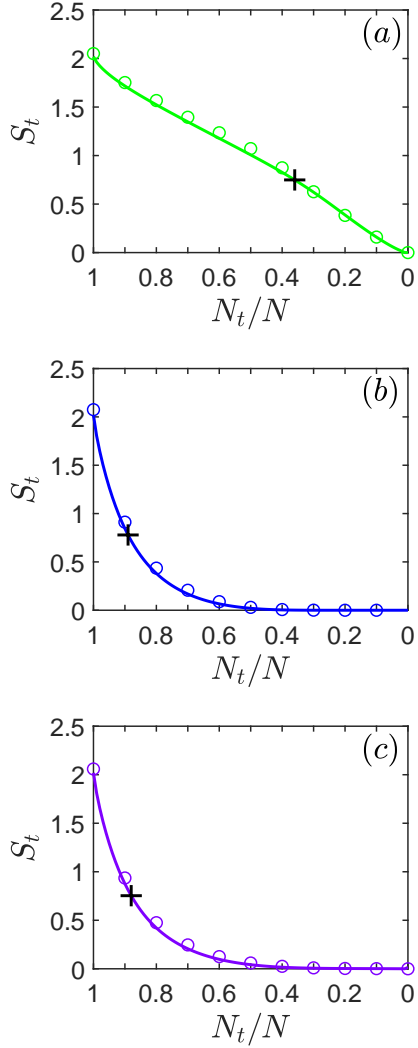}
\caption{
(Color online)
The relative entropy $S_t$ 
vs. $N_t/N$,
obtained from numerical integration of the master equation
(solid lines)
for random deletion (a), preferential deletion (b) 
and propagating deletion (c) ,
starting from a BA network with $m=5$ and size $N=10,000$.
The master equation results are in very good agreement with
the results obtained from computer simulations (circles).
The $+$ symbols mark the points at which $S_t$ decays 
to $1/e$ of its initial values.
In the case of random deletion this occurs around $N_t/N \simeq 0.4$ while in
the other two scenarios it occurs much earlier, at $N_t/N \simeq 0.9$, following the
deletion of only about $10 \%$ of the nodes.
Note that in the preferential deletion and the propagating deletion $S_t$ 
decays very quickly and practically vanishes when more than a half of the
nodes still remain.
}
\label{fig:4}
\end{center}
\end{figure}

In Fig.
\ref{fig:4}
we present the relative entropy $S_t$ 
obtained from numerical integration of the master equation
(solid lines)
for random deletion (a), preferential deletion (b) 
and propagating deletion (c),
starting from a BA network with $m=5$ and size $N=10,000$.
The master equation results are in very good agreement with
the results obtained from computer simulations (circles).
The $+$ symbols mark the points at which $S_t$ decays 
to $1/e$ of its initial values.
In the case of random deletion this occurs around $N_t/N \simeq 0.4$ while in
the other two scenarios it occurs much earlier, at $N_t/N \simeq 0.9$, following the
deletion of only about $10 \%$ of the nodes.
Note that in the preferential and the propagating deletion scenarios $S_t$ 
decays very quickly and practically vanishes when more than a half of the
nodes still remain.

\section{The degree-degree correlation function}

An important distinction in network theory is between 
uncorrelated random networks and networks 
that exhibit degree-degree correlations. 
These correlations are positive (negative) in assortative (disassortative) networks,
in which high degree nodes tend to connect to
high (low) degree nodes and low degree nodes tend to connect to 
low (high) degree nodes
\cite{Newman2002,Newman2003}.
To quantify the degree-degree correlations we define
the joint degree distribution 
$P_t(k,k')$
of pairs of nodes that reside on both sides of 
a randomly selected edge.
The marginal degree distribution, obtained by tracing over all possible 
values of $k'$, is given by

\begin{equation}
\widetilde P_t(k) = \frac{k}{ \langle K \rangle_t} P_t(k).
\end{equation}

\noindent
The degree-degree correlation function 
is given by

\begin{equation}
\Gamma_t = 
\langle K K' \rangle_t -  \langle \widetilde K \rangle_t  \langle \widetilde K \rangle_t.
\label{eq:Gamma}
\end{equation}

\noindent
The first term 
in
Eq. (\ref{eq:Gamma}) is
a mixed second moment of the form

\begin{equation}
\langle K K' \rangle_t = 
\sum_{k=1}^{\infty}
\sum_{k'=1}^{\infty}
k k' P_t(k,k'),
\end{equation}

\noindent
where the sums run over all the possible 
combinations of the degrees of pairs of adjacent nodes.
In the second term of
Eq. (\ref{eq:Gamma}), 
the mean degree $\langle \widetilde K \rangle_t$
of the degree distribution $\widetilde P_t(k)$ 
of nodes adjacent to a randomly selected edge
is given by

\begin{equation}
\langle \widetilde K \rangle_t = \sum_{k=1}^{\infty} k \widetilde P_t(k).
\end{equation}

\noindent
In case that there are no
degree-degree correlations the joint degree distribution
of pairs of adjacent nodes is given by

\begin{equation}
P_t(k,k') =  \widetilde P_t(k) \widetilde P_t(k'),
\end{equation}

\noindent
and the correlation function satisfies $\Gamma_t=0$.
This is indeed the case in configuration model networks.
However, BA networks exhibit degree-degree correlations 
and are disassortative, namely high degree
nodes tend to connect to low degree nodes and vice versa 
\cite{Fotouhia2013}.

The master equation (\ref{eq:dP/dt}) follows the time evolution
of the degree distribution $P_t(k)$ during the contraction process,
but does not account for degree-degree correlations.
Therefore, it cannot be used to explore the time dependence of
the degree-degree correlation function $\Gamma_t$.
To examine the effect of network contraction processes on
degree-degree correlations we use computer simulations.

\begin{figure}
\begin{center}
\includegraphics[width=5.4cm]{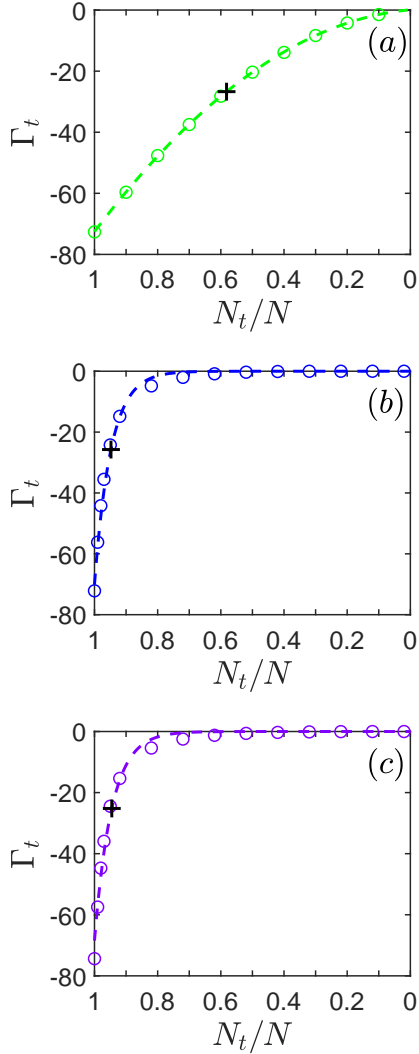}
\caption{
(Color online)
The correlation function
$\Gamma_t$
vs. $N_t/N$,
obtained from computer simulations (circles)  
for random deletion (a), preferential deletion (b) 
and propagating deletion (c),
starting from a BA network with $m=5$ and size $N=10,000$.
In the case of random deletion the simulation results  are very well fitted by
$\Gamma_t \sim (N_t/N)^2$ 
while the simulation results of the preferential deletion and the propagating
deletion processes are very well fitted by an exponential fit (dashed lines).
The $+$ symbols mark the points at which $\Gamma_t$ decays 
to $1/e$ of its initial values.
In the case of random deletion this occurs around $N_t/N \simeq 0.6$ while in
the other two scenarios it occurs at $N_t/N \simeq 0.9$, following the
deletion of only about $10 \%$ of the nodes.
Note that in the preferential deletion and the propagating deletion $\Gamma_t$ 
decays very quickly and practically vanishes when more than a half of the
nodes still remain.
}
\label{fig:5}
\end{center}
\end{figure}

In Fig.
\ref{fig:5}
we present the degree-degree
correlation function $\Gamma_t$
obtained from computer simulations (circles) of the contraction process
of BA networks of size $N=10,000$ with $m=5$,
via random deletion (a),
preferential deletion (b) and propagating deletion (c).
In the case of random deletion the simulation results are very well fitted by
$\Gamma_t \sim (N_t/N)^2$ 
while the simulation results of the preferential deletion and the propagating
deletion processes are very well fitted by an exponential fit (dashed lines).
The $+$ symbols mark the points at which $\Gamma_t$ decays 
to $1/e$ of its initial values.
In the case of random deletion this occurs around $N_t/N \simeq 0.6$ while in
the other two scenarios it occurs at $N_t/N \simeq 0.9$, following the
deletion of only about $10 \%$ of the nodes.
Note that in the preferential deletion and the propagating deletion $\Gamma_t$ 
decays very quickly and practically vanishes when more than a half of the
nodes still remain.

Putting together the results of the last two sections,
the convergence of the
degree distribution towards a Poisson distribution
(as demonstrated by the decay of $S_t$)
and the decay of the degree-degree correlations
(measured by $\Gamma_t$)
imply that networks that contract via one of the three node deletion scenarios
discussed in this paper converge towards the ER structure.

\section{Discussion}

The time scales involved in network contraction processes span over
many orders of magnitude, from fractions of a second in computer networks to  
months and years in social networks to millennia in ecological networks.
In some cases the contraction may proceed all the way down to 
the percolation threshold and into the sub-percolating regime.
In other cases only limited contraction is possible, 
either because the faulty nodes are quickly fixed,
or because the failure of a few nodes is sufficient to 
cause an unrecoverable damage to the entire system.

It is worth mentioning that there are other
network dismantling processes that involve 
optimized attacks, which maximize the damage to the network for a 
minimal set of deleted nodes
\cite{Braunstein2016}.
Such optimization is achieved by first decycling the network, 
namely by selectivly deleting nodes that reside on cycles, thus
driving the giant component into a tree structure. 
The branches of the tree are then trimmed such
that the giant component is quickly disintegrates.
Clearly, networks that are exposed to these optimized dismantling 
processes do not converge towards an ER structure.

The convergence of a contracting network towards the ER
structure takes place over a limited range of network sizes 
and densities, bounded from above by the initial size $N$
and mean degree $\langle K \rangle_0$, and from below by
the size at which the remaining network becomes fragmented
and consists of small isolated components and isolated nodes.
However, this range can be extended indefinitely by starting the contraction
process from a larger and denser network.

Network contraction processes belong to a broad
class of dynamical processes that exhibit intermediate asymptotics
\cite{Barenblatt1996,Barenblatt2003}.
The ubiquity of such processes is expressed in the following
quotation from the opening paragraph of Ref. 
\cite{Barenblatt2003}:
"In constructing the idealizations the 
phenomena under study should be considered at 'intermediate'
times and distances... 
These distances and times should be sufficiently large 
for details and features which are of secondary importance 
to the phenomenon to disappear. At the same time they 
should be sufficiently small to reveal features of the 
phenomena which are of basic value." 

\section{Summary}

In summary, we analyzed the evolution of network structure during
generic contraction processes, using the master equation, the 
relative entropy and degree-degree correlations. 
We showed that in generic contraction scenarios,
namely random, preferential and propagating deletion processes,
the network structure converges towards the
ER structure, which exhibits a Poisson degree distribution and 
no degree-degree correlations.
These results have important implications in real world situations.
For example, in cascading failures they imply that the part of the
network that continues to function  is likely to converge towards an ER structure.
In the context of ecological networks, they imply that mass extinctions
not only reduce the number of species but may also alter the structure of the
networks describing the interactions between them from  
scale-free-like networks to ER-like networks.
To conclude, while scale-free network structures with power-law degree distributions
are predominant in a world of growing or expanding networks, 
the uncorrelated Poisson-distributed ER structures are expected to be widespread
in a world of contracting networks.

This work was supported by the Israel Science Foundation grant no. 
1682/18.

\appendix

\section{Detailed derivation of the master equation}

Below we derive the master equation 
describing the temporal evolution of the degree distribution
$P_t(k)$
during network contraction via
random node deletion, preferential node deletion and propagating node deletion.

\subsection{Random node deletion} 

In the random node deletion scenario at each time
step a random node is deleted from the network together with its links.
To derive an equation for the time dependence of $N_t(k)$
one needs to account for the primary effect of the deletion 
of a node of degree $k$ and for the secondary effect in which
nodes of degrees $k$ and $k+1$ lose a link due to the deletion
of an adjacent node.
The probability that the deleted node is of degree $k$ is
given by $N_t(k)/N_t$.
Therefore, the contribution of the primary effect of node deletion to the time
derivative of $N_t(k)$ is given by
$R_t(k \rightarrow \varnothing)$ [Eqs. (\ref{eq:Rk}) and (\ref{eq:RWW})].
Regarding the secondary effect, in case that the 
node deleted at time $t$ is of degree $k'$,
it affects $k'$ other nodes, which lose one link each. 
Among these $k'$ nodes, the probability of each one of them
to be of degree $k$ is given by
$k N_t(k)/[ N_t \langle K \rangle_t ]$.
Summing up over all the possible values of the degree $k'$ of the
deleted node and evaluating the expectation value of the number of nodes
of degree $k$ that are connected to the deleted node we
obtain the secondary effect of random node deletion on nodes of degree $k$.
The rate at which nodes of degree $k$ lose one link and are reduced to degree
$k-1$ is given by

\begin{equation}
W_t(k \rightarrow k-1) =
\sum_{k'=1}^{\infty} \frac{N_{t}(k')}{N_t}
\frac{k'kN_{t}(k)}{N_t \langle K \rangle_{t}}
= \frac{k N_{t} (k)}{N_t}.
\end{equation}

\noindent
Similarly, the rate at which nodes of degree $k+1$ lose one link
and are reduced to degree $k$ is given by

\begin{equation}
W_t(k+1 \rightarrow k) =
\frac{(k+1) N_{t} (k+1)}{N_t}.
\end{equation}

\noindent
Combining the results for the primary and the secondary effects
it is found that
the time dependence of $N_t(k)$ 
is given by

\begin{equation}
\frac{d}{dt} N_t(k) 
= \frac{(k+1)}{N_t}
\left[ N_t(k+1) - N_t(k) \right].
\end{equation}

\noindent
Inserting this result into Eq. (\ref{eq:dPt_Nt})
we obtain the master equation 

\begin{equation}
\frac{d}{dt} P_t(k)=
\frac{1}{N_t}
\left[ (k+1)P_t(k+1) - k P_t(k) \right].
\label{eq:dP(t)/dtRC}
\end{equation}

\noindent
In Appendix B we present an exact solution of Eq. (\ref{eq:dP(t)/dtRC}),
which provides the time-dependent degree distribution $P_t(k)$ for any
initial degree distribution $P_0(k)$.

\subsection{Preferential node deletion}

In the scenario of
preferential node deletion, at each time step a node is selected
for deletion with probability proportional to its degree.
The probability that the node selected for deletion at time $t$ 
is of degree $k$ 
is given by 
$k N_t(k)/\left[ N_t \langle K \rangle_t \right]$.
In case that the node selected for deletion at time $t$ 
is of degree $k'$ there are $k'$ other
nodes that will be affected, losing one
link each. 
The probability of each one of these $k'$ nodes to be of degree
$k$ is given by 
$k N_t(k)/\left[N_t \langle K \rangle_t \right]$.
Summing up over all the possible values of the degree $k'$ 
and evaluating the expectation value of the number of nodes
of degree $k$ that are connected to the deleted node we
obtain that the secondary effect on nodes of degree $k$ is
given by

\begin{equation}
W_t(k \rightarrow k-1)
=
\sum_{k'=1}^{\infty}
\left[ \frac{k'N_t(k')}{N_t \langle K \rangle_t} \right]
\left[ \frac{k'k N_{t}(k)}{N_t \langle K\rangle_t} \right]
=
\frac{\langle K^2 \rangle_t}{\langle K \rangle_t^2 N_t}
kN_t(k).
\end{equation}

\noindent
Similarly, the secondary effect on nodes of degree $k+1$ is given by

\begin{equation}
W_t(k+1 \rightarrow k)
=
\frac{\langle K^2 \rangle_t}{\langle K \rangle_t^2 N_t}
(k+1) N_t(k+1).
\end{equation}

\noindent
Summing up the contributions of the primary and the secondary effects
we obtain the time derivative of $N_t(k)$,
which is thus given by 

\begin{equation}
\frac{d}{dt} N_{t}\left(k\right) 
= 
\frac{\langle K^{2}\rangle_t}{\langle K\rangle_t^{2}N_t}
\left[\left(k+1\right)N_{t}\left(k+1\right)-k N_{t}\left(k\right)\right]
-
\frac{k}{\langle K\rangle_t N_t}
N_{t}\left(k\right).
\end{equation}

\noindent
Inserting this result into Eq. (\ref{eq:dPt_Nt})
we obtain the master equation

\begin{equation}
\frac{d}{dt} P_{t}\left(k\right)
= 
\frac{\langle K^{2}\rangle_t}{\langle K\rangle_t^{2} N_t}
\left[\left(k+1\right)P_{t}\left(k+1\right) - k P_{t}\left(k\right)\right]
-
\frac{k-\langle K\rangle_t }{\langle K\rangle_t N_t} P_{t}\left(k\right).
\label{eq:dP/dtPND}
\end{equation}

\subsection{Propagating node deletion}

The propagating node deletion scenario describes network contraction
processes such as cascading failures, 
in which the damage propagates from a deleted node 
to its neighbors. 
In this scenario, at each time step we delete a random neighbor of
the node deleted in the previous step.
In case that the last deleted node does not have any yet-undeleted
neighbor, we pick a random node, select a random neighbor of this node
for deletion and continue the process from there.
The probability that the node deleted at time $t$ 
will be of degree $k'$ is
given by
$k' N_t(k')/\left[ N_t \langle K \rangle_t \right]$.
One of these $k'$ edges connects it to the node deleted
in the previous time step and another edge connects it to the node
to be deleted in the next time step.
Apart from these two neighbors, there are $k'-2$ neighbors that
lose one link each upon deletion of a node of degree $k'$.
The probability of each one of these $k'$ nodes to be of degree $k$
is given by
$k N_{t}\left(k\right)/\left[N_t \langle K \rangle_t \right]$.
Summing up over all the possible degrees $k'$ of the
node deleted at time $t$, 
we obtain the 
secondary effect on nodes of degree $k$,
which is given by

\begin{equation}
W_t(k \rightarrow k-1)
=
\frac{\langle K^{2}\rangle_t - 2 \langle K\rangle_t }{\langle K \rangle_t^2 N_t} 
k N_{t}\left(k\right).
\end{equation}

\noindent
Similarly, the secondary effect on nodes of degree $k+1$ is

\begin{equation}
W_t(k+1 \rightarrow k)
=
\frac{\langle K^{2}\rangle_t - 2 \langle K\rangle_t }{\langle K \rangle_t^2 N_t} 
(k+1) N_{t}\left(k+1\right).
\end{equation}

\noindent
The complete equation describing the time dependence of 
$N_t(k)$
is thus given by

\begin{equation}
\frac{d}{dt} N_{t}\left(k\right)
= 
\frac{\langle K^{2} \rangle_t - 2 \langle K\rangle_t }{\langle K \rangle_t^{2} N_t}
\left[\left(k+1\right)N_{t}\left(k+1\right)-kN_{t}\left(k\right)\right]
-
\frac{k}{N_t \langle K \rangle_t} N_t(k).
\end{equation}

\noindent
Inserting this result into Eq. (\ref{eq:dPt_Nt})
we obtain the master equation

\begin{equation}
\frac{d}{dt} P_t(k)
=
\frac{\langle K^2 \rangle_t - 2 \langle K \rangle_t}{\langle K \rangle_t^2N_t}
\left[ (k+1) P_t(k+1) - k P_t(k) \right]
-
\frac{k- \langle K \rangle_t}{\langle K \rangle_t N_t}
P_t(k).
\label{eq:dP(t)/dtSAW}
\end{equation}

\section{Exact solution of the master equation for random node deletion}

Below we solve Eq. (\ref{eq:dP(t)/dtRC}) for a general initial degree
distribution, given by $P_0(k)$.
To this end, we define the generating function

\begin{equation}
G(x,t) = \sum_{k=0}^{\infty} x^k P_t(k).
\end{equation}

\noindent
The initial condition of the generating function is 
denoted by 

\begin{equation}
G(x,0) = G_0(x) = \sum_{k=0}^{\infty} x^k P_0(k),
\end{equation}

\noindent
while $G(1,t)=1$ at all times due to the normalization of $P_t(k)$.
Multiplying Eq. (\ref{eq:dP(t)/dtRC}) by $x^k$ and taking a sum
over all values of $k$, we obtain the following differential equation for 
$G(x,t)$.

\begin{equation}
\frac{\partial}{\partial t} G(x,t) = 
\left( \frac{1-x}{N-t} \right)
\frac{\partial}{\partial x} G(x,t).
\label{eq:pde}
\end{equation}

\noindent
In general, the solution of Eq. (\ref{eq:pde}) must take the form

\begin{equation}
G(x,t) = f[t+(N-t)x].
\label{eq:Gf}
\end{equation}

\noindent
Inserting $t=0$ in Eq. (\ref{eq:Gf}), we find that
$f(y) = G_0(y/N)$. Therefore,

\begin{equation}
G(x,t) = G_0 \left[ \frac{t}{N} + \left(1- \frac{t}{N} \right)x \right].
\end{equation}

\noindent
Using the expression of $G_0(x)$ in terms of $P_0(k)$, we obtain

\begin{equation}
G(x,t) = \sum_{k=0}^{\infty} \left[ \frac{t}{N} + \left(1 - \frac{t}{N} \right)x \right]^k P_0(k).
\end{equation}

\noindent
Using the binomial expansion of $[t/N+(1-t/N)x]^k$, we obtain

\begin{equation}
G(x,t) = \sum_{\ell=0}^{\infty} x^{\ell} 
\left( \frac{N-t}{t} \right)^{\ell} 
\sum_{k=\ell}^{\infty} 
\binom{k}{\ell} \left( \frac{t}{N} \right)^k P_0(k).
\end{equation}

\noindent
Therefore,

\begin{equation}
P_t(k) = 
\left( 1 - \frac{t}{N} \right)^k
\sum_{k'=k}^{\infty}
\binom{k'}{k} \left( \frac{t}{N} \right)^{k'-k} 
P_0(k').
\end{equation}

\noindent
No such solution exists for the master equations describing the
preferential deletion and for the propagation deletion scenarios,
which are presented above, in Appendix A. 
Therefore, one needs to rely on numerical integration of the 
master equations.


\end{document}